\documentclass[twocolumn,
10pt,
amsmath,
amssymb,
nofootinbib,
showpacs,
superscriptaddress,
floatfix
]{revtex4-1}
\usepackage{graphicx}
\usepackage{color}
\usepackage[usenames,dvipsnames]{xcolor}
\usepackage[colorlinks=true,
linkcolor=blue,
citecolor=blue,
linktoc=page]{hyperref}
\usepackage{multirow}
\usepackage{float}
\usepackage{flushend}
\usepackage{balance}
\usepackage[varg]{txfonts}
\usepackage{ulem}
\usepackage{fancyhdr}
\usepackage{braket}
\usepackage{mathtools}
\usepackage{lipsum}
\begin{document}

\title{Heat switch and thermoelectric effects \\ based on Cooper-pair splitting and elastic cotunneling}

\author{N.\,S.\ Kirsanov}
    \affiliation{Moscow Institute of Physics and Technology, 141700, Institutskii
    Per. 9, Dolgoprudny, Moscow Distr., Russian Federation}

\author{Z.\,B.\ Tan}
    \affiliation{Low Temperature Laboratory, Department of Applied Physics, Aalto University, P.O. Box 15100, FI-00076 AALTO, Finland}

\author{D.\,S.\ Golubev}
    \affiliation{Low Temperature Laboratory, Department of Applied Physics, Aalto University, P.O. Box 15100, FI-00076 AALTO, Finland}

\author{P.\,J.\ Hakonen}
    \affiliation{Low Temperature Laboratory, Department of Applied Physics, Aalto University, P.O. Box 15100, FI-00076 AALTO, Finland}

\author{G.\,B.\ Lesovik}
    \affiliation{Moscow Institute of Physics and Technology, 141700, Institutskii
    Per. 9, Dolgoprudny, Moscow Distr., Russian Federation}

\begin{abstract}
In this paper, we demonstrate that the hybrid normal-superconducting-normal (NSN) structure has potential for a multifunctional thermal device which could serve for heat flux control and cooling of microstructures. By adopting the scattering matrix approach, we theoretically investigate thermal and electrical effects emerging in such structures due to the Cooper pair splitting (CPS) and elastic cotunneling phenomena. We show that a finite superconductor can, in principle, mediate heat flow between normal leads, and we further clarify special cases when this seems contradictory to the second law of thermodynamics. Among other things, we demonstrate that the CPS phenomenon can appear even in the simple case of a ballistic NSN structure.
\end{abstract}

\maketitle

\section{Introduction}

Superconductors are typically regarded as thermal insulators because at temperatures much less than the superconducting energy gap, $\Theta\ll\Delta$ ($\Theta$ is expressed in energy units), their thermal conductivity is exponentially small.
Nevertheless, this is not necessarily true in the case of a finite NSN structure.

\begin{figure}[t]
    \noindent\centering{
    \includegraphics[width=70mm]{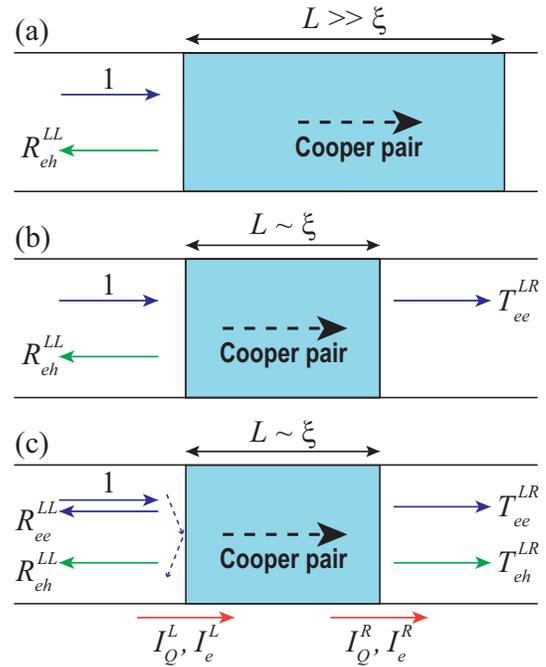}
    }
    \caption{Different NSN configurations.
    (a) The superconductor is large. An incident electron (blue) reflects as a hole (green) with probability equal to unity, $R^{LL}_{eh}=1$.
    (b) The length of the superconductor is comparable with the coherence length $\xi$. Along with AR, the EC becomes efficient: an incident electron can either propagate as an electron or reflect as a hole, $R^{LL}_{eh} + T^{LR}_{ee} = 1$.
    (c) Once normal scattering on the NS interface becomes possible, the CPS process enables electrons to be transmitted as holes. Due to the unitary condition, $R^{LL}_{ee}+R^{LL}_{eh}+T^{LR}_{ee}+T^{LR}_{eh} = 1$. Red arrows indicate the selected direction of the thermal ($I^{L(R)}_Q$) and electrical ($I^{L(R)}_e$) currents in the left and right normal leads.}
    \label{intro}
\end{figure}

Consider a ballistic NSN contact at low temperatures.
If the length of the superconducting region ($L$) is significantly larger than the superconductor coherence length ($\xi$), the quantum transport is completely defined by Andreev reflection (AR)\,\cite{Andreev::1964} (see Fig.\,\ref{intro}(a)).
Due to the fact that the subgap transport is fully determined by the Cooper pairs and the total energy of the electrons in a pair is zero (counted from the chemical potential), heat does not propagate through the superconductor.
In the case of a finite superconducting region, however, the EC process appears (Fig.\,\ref{intro}(b)), which gives rise to electrical and thermal currents through the superconductor.

Besides AR and EC, yet another process occurs in the presence of electron-to-electron scattering on the NS border. In this CPS process\,\cite{Lesovik::2001,Recher::2001} (Fig.\,\ref{intro}(c)),
two electrons from the opposite normal regions combine to form a Cooper pair.
Alternatively, one can say that an incident electron from one side is transmitted as a hole to the other side.
This phenomenon, also referred to as crossed Andreev reflection (CAR)\,\cite{Byers::1995,Deutscher::2000}, allows for the subgap energy flow and changes the thermal properties of a hybrid NSN structure.
\begin{figure*}[t]
    \noindent\centering{
    \includegraphics[width=176mm]{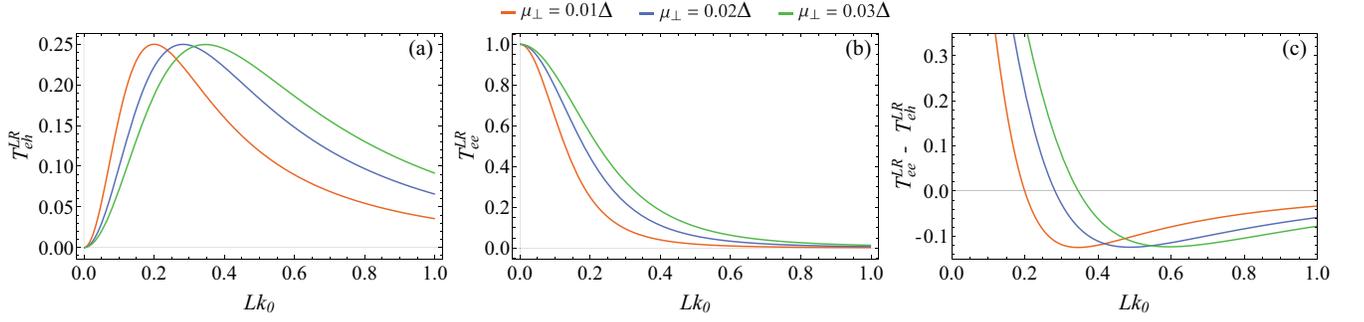}
    }
    \caption{(a) Electron-to-hole ($T^{LR}_{eh}$) and (b) electron-to-electron ($T^{LR}_{ee}$) transmission probabilities and (c) difference between transmission probabilities of EC and CPS processes as functions of the superconductor length $L$ in a ballistic NSN structure ($k_0 = \sqrt{2m\Delta/\hbar^2}$) for $\mu_\perp=\{0.01\Delta,\,0.02\Delta,\,0.03\Delta\}$ (orange, blue and green lines respectively). The plots represent the limiting case where $\varepsilon=0$ and $\mu_\perp\ll\Delta$.}
    \label{fig4}
\end{figure*}
CPS process attracts particular attention  since it potentially provides an efficient way for generating entangled electron pairs in solid state systems\,\cite{Lesovik::2001,Sadovskyy::2015,Cao::2015}; for example, by employing pair selection via well-defined energy levels in quantum dots\,\cite{Recher::2001}. Until now, CPS has been demonstrated experimentally in systems involving a superconductor connected to ferromagnetic leads\,\cite{Beckman}, to bulk normal metal leads\,\cite{Russo,CC},
to carbon nanotubes\,\cite{Strunk,Basel,Basel2}, to InAs nanowire\,\cite{Weizmann}, to self-assembled InAs quantum dots\,\cite{Tokyo} and to graphene quantum dots in Coulomb blockade regime\,\cite{Tan::2015,Tokyo2}.

Thermoelectric effects in mesocopic systems have been extensively studied in quantum dots\,\cite{Staring1993,Godijn1999,Small2003,Llaguno2004,Scheibner2005}, Andreev interferometers\,\cite{Eom2005,Jiang2005}, atomic point contacts\,\cite{Ludoph1999,Reddy2007,Widawsky2012}, and, lately, in nanowire heat engines\,\cite{Linke}.
Among other things, considerable attention has been given to the manifestations of thermoelectricity in the superconducting systems\,\cite{Hofer1,Hofer2}.
For instance, it has been predicted\,\cite{Machon} that thermoelectricity may be witnessed in ferromagnet-superconductor-based CPS devices.
To date, a growing number of papers have also examined thermoelectricity in bulk non-magnetic hybrid NSN structures by means of quasiclassical techniques based on Eilenberger and Usadel equations, see e.g. Refs.\,\cite{Tero,KZ2,KZ3}.
In particular, Cao \textit{et al.} suggested\,\cite{Cao::2015} that the CPS may occur in the sole presence of the temperature difference between the normal leads with no bias voltages applied.
In the present work, building on the scattering matrix approach, we investigate thermal and thermoelectric effects arising from CPS and EC in NSN structures going beyond quasiclassics.
We explicitly show that the superconductor can in principle mediate heat flux.
We also clarify certain cases where the CPS and EC processes seem to be in contradiction with the second law of thermodynamics.
Intriguingly, we demonstrate that CPS can occur even in the trivial case of a ballistic NSN structure.
We then consider how the CPS and EC effects can be utilized in heat transport control.
Finally, we discuss a possible experimental setting which would facilitate detection of the considered effects.

\section{NSN Thermal Properties}

Let us start by considering thermal properties of the NSN structure at low temperatures, $\Theta\ll\Delta$.
Assume that electrons in the normal parts are non-interacting.
In this case the left-to-right heat current in the left (right) normal region $I^{L(R)}_Q$ (see Fig.\,\ref{intro}(c)) is given by\,\cite{Lesovik::2011}
\begin{widetext}
\begin{align}
\label{hc1}
    I^{L(R)}_Q
    = (-) \frac{2}{h} \int d\varepsilon\bigg\{
    \Big(\varepsilon-eV_{L(R)}\Big)\,\Bigr([1 - R^{LL(RR)}_{ee}]\,f_{L(R)}
    - T^{RL(LR)}_{ee}\,f_{R(L)}\Bigr)
    +\Big(\varepsilon+eV_{L(R)}\Big)\,\Bigr(R^{LL(RR)}_{eh}\, [1 - f_{L(R)}]
    + T^{RL(LR)}_{eh}\, [1 - f_{R(L)}]\Bigr)\bigg\},
\end{align}
% \begin{align}
% \label{hc1}
%     I^{L(R)}_Q
%     = (-) \frac{2}{h} \int d\varepsilon\bigg\{
%     \Big(\varepsilon-eV_{L(R)}\Big)\,\Bigr([1 - R^{LL(RR)}_{ee}]\,f_{L(R)}&\notag\\
%     - T^{RL(LR)}_{ee}\,f_{R(L)}\Bigr)&\notag\\
%     +\Big(\varepsilon+eV_{L(R)}\Big)\,\Bigr(R^{LL(RR)}_{eh}\, [1 - f_{L(R)}]&\notag\\
%     + T^{RL(LR)}_{eh}\, [1 - f_{R(L)}]\Bigr)\bigg\}&,
% \end{align}
\end{widetext}
where $\varepsilon$ is the energy of the incident particles counted from the superconductor's chemical potential $\mu_S$; $e=-|e|$ is the charge of electron; $V_{L(R)}$ is the bias voltage of the left (right) normal lead;  $R^{LL(RR)}_{ee(eh)}$ and $T^{LR(RL)}_{ee(eh)}$ are the energy dependent probabilities that an electron incident in the left (right) lead is, respectively, reflected and transmitted as an electron (hole), see Fig.\,\ref{intro}(c); $f_{L(R)}$ is the Fermi distribution in the left (right) lead (for convenience, we omit the notation for the dependence on $\varepsilon$). 
The probabilities $T^{LR(RL)}_{ee}$ and $T^{LR(RL)}_{eh}$ correspond to EC and CPS processes respectively.
The factors $(\varepsilon\pm eV_{L(R)})$ express the fact that in general, the chemical potential is not the same for electrons and holes.
Indeed, in the presence of the bias voltage $V>0$, adding a negative electron to the reservoir requires less energy than adding a positive hole.

Now, let us consider the situation where the temperature in the left terminal is higher than in the right one and there is no voltage bias in the system, i.e., the chemical potentials are the same in all parts of the NSN junction, $\mu_L=\mu_R=\mu_S$. Assuming that $\Theta_R = \Theta_0$, $\Theta_{L} = \Theta_0 + \delta \Theta$, $\Theta_0 \gg \delta \Theta > 0$, from Eq.\,(\ref{hc1}) we obtain a non-negative left-to-right heat current:
\begin{multline}
\label{hc3}
    I^L_Q = I^R_Q = \frac{2}{h} \int d\varepsilon\, \varepsilon\, \Bigl.\frac{\partial f}{\partial \Theta}\Bigl|_{\Theta=\Theta_0} \,\delta\Theta\,[1 - R^{LL}_{eh} - R^{LL}_{ee}] \\= \frac{\delta\Theta}{2h \Theta_0^2} \int d\varepsilon\, \frac{\varepsilon^2\,[T^{LR}_{ee} + T^{LR}_{eh}]}{\cosh^2{\frac{\varepsilon}{2\Theta_0}}} > 0.
\end{multline}
Note that since the number of quasiparticles in the superconductor is exponentially small ($n^S_{e,h} \sim e^{-\Delta/\Theta_S}$), they do not contribute to the thermal current.
For $L\gg\xi$, we have $T^{LR(RL)}_{eh}=T^{LR(RL)}_{ee}\equiv0$, and the thermal current vanishes as expected.
In other situations, the thermal current may occur due to the CPS and EC processes.
% The latter could have a purely superconducting nature.
% To reflect on this assumption, one should remember that the wave function of a single electron does not exits in the superconductor.
% Therefore, we suggest that EC is nothing more than a composition of CPS and AR.

Equation\,(\ref{hc3}) remains valid even if the temperature in the superconducting region is higher than in the normal leads,  $\Theta_S > \Theta_L > \Theta_R $.
This may seem contradictory to the second law of thermodynamics, since apparently the heat flows from the colder left normal region to the warmer superconducting region.
The subject of quantum thermodynamics and quantum extension of the second law has attracted much attention lately.
It has been discovered that under certain circumstances the law in its classical sense can be violated\,\cite{Lesovik::2016,Lebedev::2016,Kirsanov::2018(1),Kirsanov::2018(2)}.
% However, in the present case one must not forget that the left lead and the superconductor do not constitute an isolated system.
% With a view to this fact, it can be argued that the process does not violate the second law.
However, in the present case the transfer of a particle from the warmer left reservoir to the colder right one is still associated with the overall increase in entropy, $\Delta S=-\frac{\varepsilon}{\Theta_L} + \frac{\varepsilon}{\Theta_R} > 0$, in which sense it does not violate the second law.
At the same time, the entropy change is non-local, and this effect may be considered non-trivial as it cannot be found in normal metal structures.
We may also note that such non-locality disappears at temperatures large compared with $\Delta$, when the transport is no longer determined by the Cooper pairs. 
%Continuing this line of argument, we may expect that at temperatures $\sim\Delta$ the effect can manifest itself only partially.

\section{NS Scattering Beyond Andreev Approximation}
\begin{figure*}[t]
    \noindent\centering{
    \includegraphics[width=170mm]{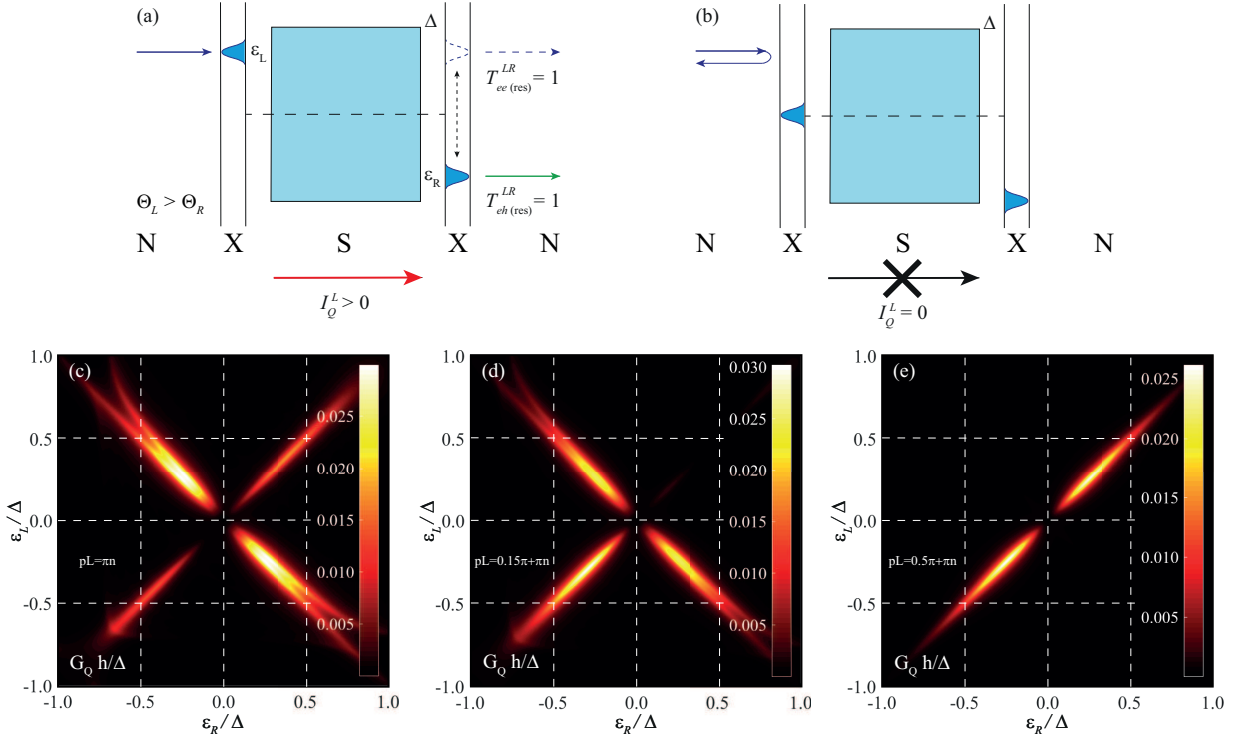}
    }
    \caption{Working principle of the NSN heat switch utilizing double barrier scatterers with the energy-dependent transmission probability (X).
    Each barrier is perfectly transparent at its resonance energy, whereas particles with other energies are completely reflected.
    The positions of the resonances can be adjusted by the external gate voltages.
    (a) The resonance settings realizing ideal transmission.
    Symmetric configuration allows for unity electron-to-hole transmission probability at resonance, $T^{LR}_{eh\,\text{(res)}}=1$; dashed resonance curve marks perfect EC configuration, $T^{LR}_{ee\,\text{(res)}}=1$.
    Heat is transferred through the system.
    (b) An example of a cutoff configuration: particles do not propagate through the switch. Heat flow is absent.
    (c,d,e) Color plots of the heat conductance, $G_Q =I^{L}_Q/\delta\Theta$, as function of the resonance energies at the left ($\varepsilon_L$) and right ($\varepsilon_R$) double barriers for $pL=\{\pi n,\, 0.15\pi + \pi n,\, 0.5\pi + \pi n\}$ ($n$ is integer). $E_F\gg\Delta$, $L=\xi_0=k_F\Delta/(2E_F)$ and $\Theta=0.1\Delta$; the parameters of the X-barriers are  $\Gamma_X=0.03\,\Delta$, $T_\text{(res)}^X=1$.
    }

    \label{heat_switch}
\end{figure*}
%
% The transport phenomena on NS border have brought much attention\,\cite{Lesovik::1998}.

In this section we discuss the situation which demonstrates that the CPS process can be observable even in a very simple system -- a finite, fully ballistic NSN structure.
% %edit
% Consequently, one may expect the appearance of the thermoelectric effects which will be discussed in Sec.\,\ref{sec5}.
To begin, it is important to recognize the conditions necessary for the CPS process to occur in the first place.

Consider a structure for which the length of the superconductor far exceeds the coherence length, i.e., the size of a Cooper pair (see Fig.\ref{intro}(a)).
In such a case, the CPS becomes highly improbable.
Instead, the paired electrons may split only in accordance with a local AR process, which forces both resulting electrons into the same normal lead.
A different NSN configuration with a finite superconducting region, as shown in Fig.\ref{intro}(b), does not necessarily constitute a CPS device either.
Nevertheless, now, as it appears from the boundary conditions, the CPS may occur if there is a non-zero probability of the electron-to-electron scattering on the NS interface (see Fig.\ref{intro}(c)).
Dzhikaev\,\cite{Dzhikaev} showed that such specular reflections can, in fact, take place, if the incident electrons move nearly parallel to the interface; several basic effects emerging from this phenomenon have been studied in Refs.\,\cite{Kadigrobov1,Kadigrobov2,Kadigrobov3}.
Here we shall explicitly show that the particles "sliding" along the interface can give rise to the CPS process.
Namely, we demonstrate that the CPS is possible with a small effective chemical potential $\mu_\perp = \mu - \hbar^2k_\parallel^2/(2m)$, where $k_\parallel$ is the wave vector's component parallel to the interface.
The value of $\mu_\perp$ depends on the electrons' angle of incidence and thus can be controlled.
Our results go beyond the Andreev approximation, in which an incident electron and a reflected hole move nearly perpendicular to the NS boundary with the wave vectors close to the Fermi wave vector $k_F$, and cannot be captured by Eilenberger equation.
The latter predicts vanishing CPS probability for a short NSN structure with fully transmitting NS boundaries\,\cite{KZ1}.
We may note that the advantage of the scattering matrix approach over the quasiclassical description is seen even in the case of the ideal NS boundary: the Eilenberger equation predicts unity electron-to-hole reflection probability, while in our framework its value can be less than one.

To conduct our analysis, we shall consider the exact solutions of the Bogoliubov--de Gennes equations\,\cite{Bogoliubov::1958,deGennes::1966} in a NSN hybrid structure. The perpendicular component of the electron's (hole's) wave vector is defined by $\hbar^2k_{+(-)}^2/(2m)=\mu_\perp \pm \varepsilon$.
In the left normal region, the two-component wave function, describing electrons ($u$) and holes ($v$), is given by
\begin{align}
\label{wfL}
    \begin{pmatrix} u \\ v \end{pmatrix} =
    &\begin{pmatrix} e^{ik_+x} + r_{ee}\,e^{-ik_+x}\\ r_{eh}\,e^{ik_-x} \end{pmatrix};
\end{align}
for the transmitted wave we have
\begin{align}
\label{wfR}
    \begin{pmatrix} u \\ v \end{pmatrix} =
    &\begin{pmatrix} t_{ee}\,e^{ik_+x}\\ t_{eh}\,e^{-ik_-x} \end{pmatrix}.
\end{align}
Here $t_{ee(eh)}$ and $r_{ee(eh)}$ are the electron-to-electron (electron-to-hole) transmission and reflection amplitudes, respectively.
In the superconducting region, the wave function is given by
\begin{align}
\label{wfS}
    \begin{pmatrix} u \\ v \end{pmatrix} =
    &\begin{pmatrix} e^{i\alpha} \\ 1 \end{pmatrix}\,\Bigr(A\,e^{ipx-qx} + B\,e^{-ipx+qx}\Bigr)\notag\\+
    &\begin{pmatrix} e^{-i\alpha} \\ 1 \end{pmatrix}\,\Bigr(C\,e^{ipx+qx} + D\,e^{-ipx-qx}\Bigr),
\end{align}
where $\alpha=\arccos{\varepsilon/\Delta}$; $p$ and $q$ are defined as $p^2-q^2=2m\mu_\perp/\hbar^2$ and $2pq=(2m\Delta/\hbar^2)\sin \alpha$.

The transmission probabilities $T^{LR}_{ee(eh)}=|t_{ee(eh)}|^2$ can be found from eight relations for the wave function's boundary conditions.
Here, we only address the limiting situation where $\varepsilon\ll\mu_\perp\ll\Delta$.
In this case, $T^{LR}_{ee}$ and $T^{LR}_{eh}$ can be calculated analytically, but the corresponding expressions become too cumbersome (see Appendix \ref{Appendix1}).
Therefore, we shall base our analysis on the numerically evaluated plots.

The dependence of $T^{LR}_{eh}$ and $T^{LR}_{ee}$ on the dimensionless parameter $L k_0$ ($k_0 = \sqrt{2m\Delta/\hbar^2}$) is shown in Fig.\,\ref{fig4}(a,b), where we choose $\mu_\perp=\{0.01\Delta,\,0.02\Delta,\,0.03\Delta\}$ and $\varepsilon=0$.
One can see that the maximum value of $T^{LR}_{eh}$ is close to $0.25$, which makes the effect quite significant.
Moreover, it should be noted that in a certain range of $L$, $T^{LR}_{eh}$ exceeds $T^{LR}_{ee}$ (see Fig.\,\ref{fig4}(c)); in other words, CPS process is stronger than EC.
Yet, for the effect to appear, $L$ should be comparable with a rather small effective coherence length $\tilde\xi=\Bigl.\frac{1}{q}\Bigl|_{\varepsilon=0}$:
\begin{equation}
\label{Lsim}
    L \sim \frac{1}{k_0} \sim \tilde\xi.
\end{equation}
In the case of aluminum superconductor, $L$ should be $\sim 10\,$nm.
We should emphasize, however, that the result is obtained for a one-dimensional structure, and may be invalid for other geometries (see discussion in Ref.\,\cite{Flensberg}).
For instance, as was pointed out in Ref.\,\cite{Recher::2001}, in the case of a three-dimensional junction, the CPS effect is suppressed if $L$ is large compared to $k^{-1}_F$, which, for metals, is typically $\sim \mathrm{\AA}$.

\section{Heat Switch}

In this section we discuss the possibility to utilize NSN structures in the control of heat transport.
Let us consider Eq.\,(\ref{hc3}) which indicates that the thermal conductivity of the structure is directly dependent on the transmission probabilities $T^{LR}_{ee}$ and $T^{LR}_{eh}$.
Therefore, if one can control these values, the structure may be operated as a heat switch\,\cite{heatvalve}, i.e., a device that switches on demand between the thermal conductor and thermal insulator modes.

As we have seen in the previous section, the electron-to-hole transport can take place even in ballistic NSN structures and furthermore, can to some extent be controlled.
In reality, however, this approach may be unsuitable for the practical needs.
To this end, we devise our heat switch using an NXSXN structure that utilizes scatterers (X) with the energy-dependent transmission probability, e.g., quantum dots\,\cite{Yeyati::2011}. The transparency function $T^X(\varepsilon)$ of an individual scatterer is characterized by its resonance position $\varepsilon_X$, resonance
half-width $\Gamma_X$ and the peak transmission probability $T_\text{(res)}^X$:
\begin{equation}
\label{lorenz}
    T^X(\varepsilon) = \frac{\Gamma_X^2\,T_\text{(res)}^X}{(\varepsilon-\varepsilon_X)^2+\Gamma_X^2}.
\end{equation}

The physics of the device we propose is based on the scattering matrices outlined in Ref.\,\cite{Sadovskyy::2015}.
From the expressions for the transmission probabilities (see Appendix \ref{Appendix2}) it follows that in the case of the symmetric resonance configuration (see Fig.\,\ref{heat_switch}(a)), the maximal electron-to-hole transmission probability, $T_{eh\,\text{(res)}}^{LR}$, can reach unity (alternatively, the resonances may be positioned on the same level; in this case, CPS would be replaced by EC and $T_{ee\,\text{(res)}}^{LR}=1$).
Conversely, certain settings may completely block the transmission; an example is depicted in Fig.\,\ref{heat_switch}(b).

A fuller understanding of the thermal properties of NXSXN structures can be achieved by considering the dependence of the heat conductance, $G_Q =I^{L}_Q/\delta\Theta$, on the positions of the left and right resonances.
The situation where the Fermi energy $E_F$ is much larger than $\Delta$ is shown in Figs.\,\ref{heat_switch}(c,d,e) (the parameters are given in the caption) plotted for $pL=\{\pi n,\, 0.15\pi + \pi n,\, 0.5\pi + \pi n\}$, where $n$ is integer (if $E_F \gg \Delta$, $pL$ can be regarded independent from $\varepsilon$).
One can see how the variation of the quantum dot gate potentials can drastically change the thermal conductivity of the system.
Further in this paper we demonstrate that it is also possible to configure the structure in such a way that it would essentially become an electrical insulator, but would still conduct thermal current.

\section{EC \& CPS Cooling}
\begin{figure}[t]
    \noindent\centering{
    \includegraphics[width=60mm]{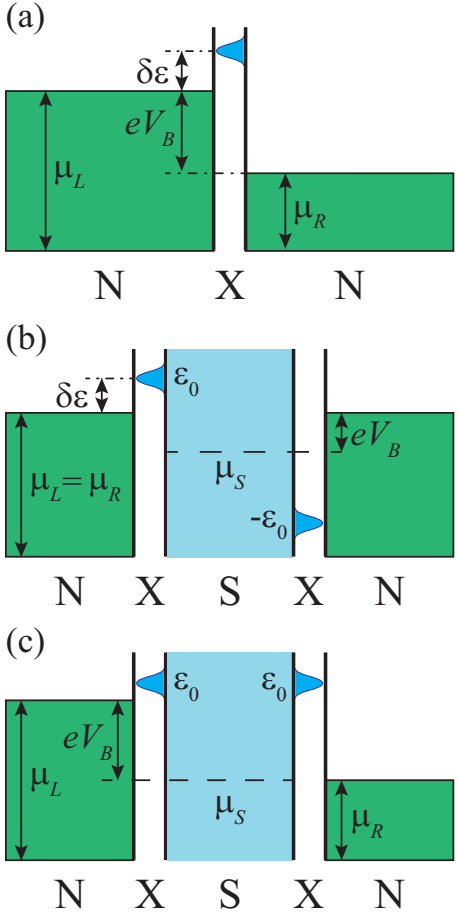}
    }
    \caption{(a) NXN refrigerator: the heat extraction from the left lead occurs due to the electron concentration difference at energies around the resonance of the X scatterer. (b) CPS refrigerator: the resonance configuration of the barriers corresponds to the perfect electron-to-hole transmission probability. Similar to the previous scheme, the cooling is based on the concentration difference of the quasiparticles in the normal leads. (c) The NXSXN configuration for which CPS is replaced by EC. The scheme is essentially reduced to the NXN one.}
    \label{cooling}
\end{figure}
In this section we discuss a CPS-based cooling device involving voltage bias.
The working principle of such a device is not distinctive to structures with superconductors; however, the presence of a superconducting electrode in some situations allows for better efficiency.
We start by discussing the NXN design\,\cite{Humphrey} and then proceed with the NXSXN version.

\subsection{NXN Scheme}
\begin{figure}[t]
    \noindent\centering{
    \includegraphics[width=85mm]{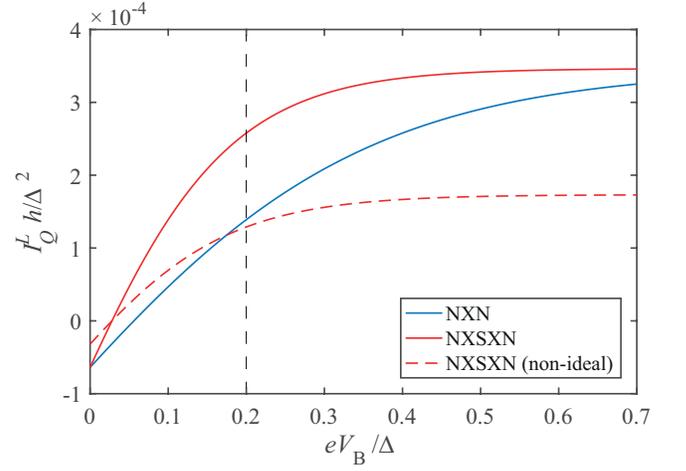}
    }
    \caption{Heat currents $I^\text{NN}_Q$ (blue;  $T^X_{\text{(res)}}=1$) and $I^\text{NSN}_Q$ (red; solid and dashed lines correspond to, respectively, $T_{eh\,\text{(res)}}^{LR}=1$ and $T_{eh\,\text{(res)}}^{LR}=0.5$ (non-ideal CPS)) as functions of the bias voltage $V_\text{B}$.
    The dashed vertical line indicates where the solid lines are separated the most. 
    The positiveness of the heat current directed from the left colder region to the right warmer one ($\Theta_R>\Theta_L$) means that the heat flows against the temperature gradient.
    The parameters are $\delta\varepsilon=0.05\Delta$, $\Theta_L=0.1\Delta$, $\Theta_R=0.2\Delta$, $\Gamma_{X}=\Gamma_{XSX}=0.003\Delta$.}
    \label{nsn_v_nn}
\end{figure}

Let us consider two normal leads connected via a quantum dot with a narrow resonance.
%minus
Suppose also that the leads are biased at negative constant voltage $V_\text{B}$, i.e., $\mu_L=\mu_R+eV_\text{B}$, and the resonance of the dot is positioned slightly above the chemical potential of the left lead, $\delta\varepsilon = \varepsilon_0-eV_\text{B} \ll eV_\text{B}$ (see Fig.\,\ref{cooling}(a)).
According to Eq.\,(\ref{hc1}), in which we put $T^{LR(RL)}_{eh}=R^{LL(RR)}_{eh}\equiv0$ and $T^{LR(RL)}_{ee}=T^X$, the left-to-right heat currents are given by
\begin{equation}
\label{NN_hc}
    I^{L(R)}_Q = \frac{2}{h} \int d\varepsilon\,(\varepsilon-\mu_{L(R)})\,T^X\,[f_L - f_R].
\end{equation}
%
%check
Supposing that the resonance is narrow, i.e., $\Gamma_X \ll \Theta_{L,R}$, the relation for the heat extraction from the left region can be rewritten as
\begin{multline}
\label{NN_cool}
    I^\text{NN}_Q \equiv I^L_Q \simeq \frac{2\,\pi\,T^X_\text{(res)}\,\Gamma_X\,\delta\varepsilon}{h}\,\Bigl[\frac{1}{e^{\delta\varepsilon/\Theta_L} + 1} - \frac{1}{e^{(\delta\varepsilon+eV_\text{B})/\Theta_R} + 1}\Bigl].
\end{multline}
We notice that the current is positive and hence the left region is cooling when the left term in the brackets is greater than the right one, i.e., when the number of electrons with energies close to $\delta\varepsilon$ (counted from $\mu_L$) in the left terminal is higher than in the right one.
The heat current may appear even opposed to the temperature gradient $\delta\Theta$ (see Fig.\,\ref{nsn_v_nn}).
Note that in its physical sense, this process is similar to Peltier effect.

\subsection{NXSXN scheme}

The NXSXN cooling device, depicted in Fig.\,\ref{cooling}(b), is based on CPS process.
Suppose, the voltage is applied in such a way that the normal leads have equal chemical potentials which are higher than that of the superconductor, $\mu_L = \mu_R = \mu_S + eV_\text{B}$.
We consider a symmetrical resonance configuration when the left resonance is situated at $\varepsilon_L=\varepsilon_0$ above the superconductor's chemical potential and the right one lies below it at $\varepsilon_R=-\varepsilon_0$ (the energies are counted from $\mu_S$).
% We also assume that such a scheme enables perfect electron-to-hole transmission, $T_{eh\,\text{(res)}}^{LR} = 1$.
Using Eq.\,(\ref{hc1}), one can find the left-to-right heat current in the left normal lead in the case where $T_{eh}^{LR}$ has a small resonance half-width $\Gamma_{XSX}\ll \Theta_{L,R}$:
\begin{multline}
\label{NSN_cool}
    I^\text{NSN}_Q \equiv I^L_Q
    = \frac{2}{h} \int d\varepsilon\,(\varepsilon-e V_\text{B})\,T^{LR}_{eh}\,\\
    \times\Bigl[\frac{1}{e^{(\varepsilon-eV_\text{B})/\Theta_L} + 1}
    - \frac{1}{e^{(\varepsilon+eV_\text{B})/\Theta_R} + 1}\Bigl]\\
    \simeq \frac{2\,\pi\,T_{eh\,\text{(res)}}^{LR}\,\Gamma_{XSX}\,\delta\varepsilon}{h}\,\Bigl[\frac{1}{e^{\delta\varepsilon/\Theta_L} + 1} - \frac{1}{e^{(\delta\varepsilon+2eV_\text{B})/\Theta_R} + 1}\Bigl],
\end{multline}
It can be seen that the heat extraction from the left region takes place when the number of electrons with energies $~\varepsilon_0$ in the left lead surpasses the number of holes at energies $~-\varepsilon_0$ in the right lead.
Thus, the operating principle of both cooling schemes is essentially the same.

The efficiency coefficient of the NXSXN refrigerator, extracting heat $I_Q^L$ from the colder left reservoir using input electrical power $W$, is given by
\begin{equation}
    \eta=\frac{I_Q^L}{W}.
\end{equation}
Here $W$ can be expressed in terms of the bias voltage and the electrical current in the left lead $I_e^L$, which will be given below: $W=I_Q^R-I_Q^L=2I_e^L V_\text{B}$. 
For $\Gamma_{XSX}\ll \Theta_{L,R}$ we have $\eta=(\varepsilon_0-e V_\text{B})/(2e V_\text{B})$.
In the limit $\varepsilon_0 \rightarrow e V_\text{B}\frac{\Theta_R+\Theta_L}{\Theta_R-\Theta_L}-0$ the efficiency assumes the Carnot value, $\eta_C=\frac{\Theta_L}{\Theta_R-\Theta_L}.$

\subsection{Advantage of CPS process}
A comparison of Eqs.\,(\ref{NN_cool}) and (\ref{NSN_cool}) shows that, provided $\Gamma_X=\Gamma_{XSX}=\Gamma$ and $T^X_{\text{(res)}}=T^{LR}_{eh\,\text{(res)}}=T_{\text{(res)}}$, at the same voltage bias and with fixed $\delta\varepsilon$, the CPS device has larger cooling power than the NXN system.
In Fig.\,\ref{nsn_v_nn} we plot $I^\text{NN}_Q$ (blue solid line) and $I^\text{NSN}_Q$ (red solid line) corresponding to the unity maximal transmission probabilities as functions of the bias voltage (the parameters are given in the caption).
The dashed vertical line marks the point where these heat currents differ the most.
If $eV_\text{B}\gg\Theta_R$, the enhancement in the heat current is close to zero:
\begin{align}
    \Delta I_Q &= I_Q^\text{NSN} - I_Q^\text{NN} \notag \\
    &\simeq \frac{2\,\pi\,T_{\text{(res)}}\,\Gamma\,\delta\varepsilon}{h}\,e^{-(\delta\varepsilon+eV_\text{B})/\Theta_R}\,\Bigl[1 - e^{-eV_\text{B}/\Theta_R}\Bigl].
\end{align}
In the real experiment it is, of course, possible that $T_{eh\,\text{(res)}}^{LR}\Gamma_{XSX} < T_{\,\text{(res)}}^{X}\Gamma_{X}$, in which case the NXN system may have an advantage at some $V_\text{B}$.
This is reflected in Fig.\,\ref{nsn_v_nn} by the red dashed line which corresponds to NXSXN cooling power for $T_{eh\,\text{(res)}}^{LR}=0.5$.

When configured as depicted in Fig.\,\ref{cooling}(c), the NXSXN scheme is essentially reduced to the NXN one. With these settings the CPS process is completely replaced by EC.

\section{Thermoelectricity and Joule heating}
\label{sec5}
We proceed by addressing thermoelectric properties of the NSN structure.	
We may write the left-to-right electric current in NSN structure in the form similar to Eq.\,(\ref{hc1})\,\cite{Lesovik::2011}:
%minus
\begin{widetext}
\begin{align}
\label{ec1}
I^{L(R)}_e = (-)\frac{2e}{h} \int d\varepsilon\, \{-R^{LL(RR)}_{eh}\, [1 - f_{L(R)}]-T^{RL(LR)}_{eh}\, [1 - f_{R(L)}]+[1 - R^{LL(RR)}_{ee}]\,f_{L(R)}-T^{RL(LR)}_{ee}\, f_{R(L)}\}.
\end{align}
\end{widetext}
% \begin{align}
% \label{ec1}
% I^{L(R)}_e = (-)\frac{2e}{h} \int d\varepsilon\, \{-&R^{LL(RR)}_{eh}\, [1 - f_{L(R)}]\notag
% \\
% -&T^{RL(LR)}_{eh}\, [1 - f_{R(L)}]\notag
% \\+&[1 - R^{LL(RR)}_{ee}]\,f_{L(R)}\notag
% \\-&T^{RL(LR)}_{ee}\, f_{R(L)}\}.
% \end{align}
%???
Bearing in mind that the currents vanish in equilibrium, at zero bias voltage this formula gives the simple result
%minus
\begin{equation}
\label{ec3}
I^R_e = \frac{2e}{h} \int d\varepsilon\, [T^{LR}_{ee} - T^{LR}_{eh}] \,\bigr(f_L-f_R).
\end{equation}
As distinct from the thermal current given by Eq.\,(\ref{hc3}), the electric current in the right lead is zero if $T^{LR}_{eh} \equiv T^{LR}_{ee}$.
This means that the NSN structure can be configured in such a way that it would conduct heat, but not electric charge.
The system therefore does not satisfy the Wiedemann--Franz (WF) law\,\cite{Franz::1853} stating that the ratio of the thermal conductivity ($\kappa$) to the electrical conductivity ($\sigma$) is proportional to the temperature, $\kappa/\sigma = L \, \Theta$
where the Lorenz number $L=\frac{\pi^2}{3}\left(\frac{k_B}{e}\right)^2$.
Previously, it has been shown that the WF law can be violated when the thermoelectric effect is significant\,\cite{Engquist::1981}.
In the case of normal metal structures, thermoelectricity is typically associated with the energy dependent transmission\,\cite{Lesovick::1989}; namely, if this dependence is weak, the effect can be expressed by the Cutler--Mott formula.
Intriguingly, this is not so, when we consider NS junctions.
Yet, the mechanisms by which the thermoelectricity can be established, may appear in the case of the finite superconductor.

Let us now suppose that the leads are biased at voltages $V_L$ and $V_R$ with respect to the superconductor.
The heating power of the structure $\dot{Q}$ can be written, using Eqs.\,(\ref{hc1}) and (\ref{ec1}), as
\begin{align}
    \dot{Q} = I^R_Q-I^L_Q = I^L_e\,V_L - I^R_e\,V_R,
\end{align}
meaning that the system obeys the Joule law.
From Eq.\,(\ref{hc1}) it can also be seen that the particles dissipate energy through relaxation to the local chemical potential.
Consequently, in contrast with the classical picture, the heating is non-homogeneous, as it can vary substantially from one part of the structure to another.
This creates a temperature gradient which, in turn, can result in thermoelectricity.

\section{Experimental considerations}
Let us explore experimental detectability for heat current caused by non-local thermal and thermoelectric effects.
A promising system is a graphene-based setting with two quantum dots etched out of exfoliated graphene \cite{Tan::2015}. We aim at non-local thermal phenomena at moderate charge density, and consequently we may neglect the inherent peculiar properties of Andreev reflection in graphene \cite{Beenakker2006,Beenakker2008,Cayssol2008}.
The advantage of graphene for Cooper pair splitting is that its electrons are quite well isolated from lattice so that a small heat input can raise the electronic temperature substantially.
Furthermore, it is quite easy to pattern part of the very same graphene flake to obtain proximity-induced superconductivity \cite{heersche2007}, which can be employed for thermometry based on switching supercurrents.
The switching current $I_\text{SW}$ of an diffusive graphene superconductor-graphene-superconductor (SGS) junction depends strongly on temperature when the Thouless energy $E_\text{Th} \ll \Delta$, which can be reached in junctions of length $L \sim 400\,\text{nm}$ for standard graphene devices on $\text{SiO}_2$\,\cite{voutilainen2010}.
However, as found out in Ref.\,\cite{voutilainen2010}, non-equilibrium quasiparticles may contribute to the heat flow out of the SGS junction and increase the coupling of the graphene sheet to the environment.
Consequently, we base our estimates on the experimental results on heat relaxation in graphene obtained in Ref.\,\cite{voutilainen2010}.
In fact, their device has dimensions and characteristics close to such a temperature detector that could be adopted for thermometry on a graphene Cooper pair splitter.

The sensitivity of a switching current thermometer depends on the width of the switching distribution and the steepness of $I_\text{SW}(\Theta_e)$.
We assume 400-nm-long SGS junctions, for which $\mathrm{d}I_\text{SW}(\Theta_e)/\mathrm{d}\Theta_e = 100 \div 200\,\text{nA/K}$\,\cite{voutilainen2010}.
At temperatures below 100 mK, we estimate for the single-measurement temperature resolution $\Delta \Theta = 3\,\text{mK}$, which can be improved to $\sim 1\,\text{mK}$ by averaging.
In the non-hysteretic regime above $450\,\text{mK}$, the temperature resolution degrades and we estimate $\Delta \Theta \simeq 10\,\text{mK}$ at $500\,\text{mK}$.

According to Ref.\,\cite{voutilainen2010}, a heating power of 150\,fW, 2.3\,pW and 9\,pW will increase the temperature of the graphene thermometer and the attached graphene heat link to about 35\,mK, 120\,mK and 190\,mK, respectively.
Using the parameter values employed in Fig.\,\ref{nsn_v_nn}, the heat current amounts to 13\,fW at bias $V_\text{B}=0.2 \,\Delta/e$.
The corresponding $\Delta \Theta \sim 3\,\text{mK}$ in the SGS detector will be detectable experimentally, although galvanic coupling between the SGS detector and the CPS structure requires careful tracking of the inadvertent current paths in the circuit.
The prospects for heat current detection, however, become much more favorable in the situation where $\Theta_L=\Theta_R=0.2\,\Delta$, $\Gamma \simeq 0.1\,\Delta$ and $\delta\varepsilon=0.2\,\Delta$, for which we can no longer apply the approximation made in Eqs.\,(\ref{NN_cool}) and (\ref{NSN_cool}), and the transmission probability shall be considered as a Lorentz function.
In this case, the heat current increases by two orders of magnitude.
According to the experimental work of Ref.\,\cite{Inverse}, the heat current may also increase by inverse proximity effect in this regime with substantial coupling between N and S conductors.
% Finally, we note that even without any optimization
% the expected value of the non-local heat flux for the sample used in the experiment \cite{Tan::2015}
% is of the order of 10 aW, see Fig. \ref{Fig::IQe}(b).
% This power is already detectable with precise on-chip thermometers, see e.g. Refs. \cite{Dutta,Alberto}.

\section{Summary}
To sum up, we have shown that the hybrid NSN structures can have promising applications in thermal regulation; namely, we have presented the concepts of the NSN-based heat switch and refrigerator.
Using scattering matrix framework, we have uncovered thermal phenomena appearing in NSN structures.
Our analytic results indicate that the heat can be conducted non-locally through a superconducting lead in the presence of the CPS and EC.
Intriguingly, we have shown that the CPS process may be witnessed even in ballistic NSN structures.
Moreover, we have addressed thermoelectricity and the Joule law manifestation in the NSN systems.
Lastly, we have made suggestions regarding the experimental detectability of the non-local effects above.

\section*{Acknowledgements}
The authors thank C. Flindt and P. Burset for fruitful discussions.
We are indebted to I.\,A. Sadovskyy for helping with numerical analysis.
This work was supported by Aalto University School of Science Visiting Professor grant to G.B.L., as well as by Academy of Finland projects 290346 (Z.B.T., AF post doc), 314448 (BOLOSE) and 312295 (CoE, Quantum Technology Finland). This research project utilized the Aalto University OtaNano/LTL infrastructure.
The work of N.S.K. and G.B.L. was supported by the Government of the Russian Federation
(Agreement № 05.Y09.21.0018), by the RFBR Grants No. 17-02-00396A and 18-02-00642A, Foundation for the Advancement of Theoretical Physics and Mathematics "BASIS",
the Ministry of Education and Science of the Russian Federation
16.7162.2017/8.9.

\textit{Note added.}-- While preparing this paper, we became aware of two related works\,\cite{Sanchez::2018,Hussein::2018}, in which the case of strong Coulomb interaction on the dots is considered.

\appendix
\section{NSN Boundary Conditions}
\label{Appendix1}
The boundary conditions for the wave functions expressed in Eqs.\,(\ref{wfL}), (\ref{wfR}) and (\ref{wfS}) can be written as
\begin{widetext}
\begin{gather}
    1+r_{ee}=e^{i \alpha}\,[A+B]+e^{-i \alpha}\,[C+D];\\
    i k_+-i k_+ r_{ee} = e^{i \alpha}\,[A\,(i p-q)+B\,(-i p +q)]+e^{-i \alpha}\,[C\,(i p + q) + D\,(-i p -q)];\\
    e^{i\alpha}\,[A\,e^{(i p - q) \,L}+ B\,e^{(-i p + q) \,L}]+e^{-i\alpha}\,[C\,e^{(i p +q) \,L} + D\,e^{(-ip-q) \,L}] = \tilde{t}_{ee};\\
    e^{i\alpha}[A\,(i p - q) \,e^{(i p - q) \,L}+ B\,(-i p + q) \,e^{(-i p + q) \,L}]+e^{-i \alpha} [C\,(i p +q) \,e^{(i p +q) \,L} + D\,(-ip-q) \,e^{(-ip-q) \,L}]=  i k_+ \tilde{t}_{ee};\\
    r_{eh} = A + B + C + D;\\
    i k_- r_{eh} = A\,(i p -q) +B\,(-i p +q) + C\,(ip + q) + D\,(-ip-q);\\
    A\,e^{(i p - q) \,L}+ B\,e^{(-i p + q) \,L}+C\, e^{(i p +q) \,L} + D\,e^{(-ip-q) \,L} = \tilde{t}_{eh};\\
    A\,(i p - q)\,e^{(i p - q) \,L}+ B\,(-i p + q)\,e^{(-i p + q) \,L}+C\,(i p +q)\,e^{(i p +q) \,L} + D\, (-ip-q)\,e^{(-ip-q) \,L} = -i k_- \tilde{t}_{eh};
\end{gather}
\end{widetext}
where $\tilde{t}_{ee(eh)} = e^{\pm ik_{+(-)}}t_{ee(eh)}$.
Let us for convenience denote $\tilde\mu_\perp=\mu_\perp/\Delta$.
In the limit $\varepsilon\ll\mu_\perp\ll\Delta$, where $p = k_0/\sqrt{2} + O\,(\mu_\perp)$, $q = k_0/\sqrt{2} + O\,(\mu_\perp)$ and $k_\pm = k_0\, \sqrt{\tilde\mu_\perp} +O\,(\varepsilon/\sqrt{\mu_\perp})$, the boundary conditions become
\begin{widetext}
\begin{gather}
    1+r_{ee}=i\,[A+B]-i\,[C+D];\\
    \sqrt{2\tilde\mu_\perp}(1- r_{ee}) = A\,(i -1)+B\,(-i +1)-C\,(i + 1) - D\,(-i -1);\\
    i\,[A\,e^{(i - 1) \,Lk_0/\sqrt{2}}+ B\,e^{(-i  + 1) \,Lk_0/\sqrt{2}}]-i\,[C\,e^{(i  +1) \,Lk_0/\sqrt{2}} + D\,e^{(-i-1) \,Lk_0/\sqrt{2}}] = \tilde{t}_{ee};\\
    A\,(i  - 1) \,e^{(i  - 1) \,Lk_0/\sqrt{2}}+ B\,(-i  + 1) \,e^{(-i  + 1) \,Lk_0/\sqrt{2}}-C\,(i  +1) \,e^{(i  +1) \,Lk_0/\sqrt{2}} - D\,(-i-1) \,e^{(-i-1) \,Lk_0/\sqrt{2}}= \sqrt{2\tilde\mu_\perp} \tilde{t}_{ee};\\
    r_{eh} = A + B + C + D;\\
    i r_{eh} \sqrt{2\tilde\mu_\perp} = A\,(i - 1) +B\,(-i + 1) + C\,(i + 1) + D\,(-i - 1);\\
    A\,e^{(i  - 1) \,Lk_0/\sqrt{2}}+ B\,e^{(-i  + 1) \,Lk_0/\sqrt{2}}+C\, e^{(i  +1) \,Lk_0/\sqrt{2}} + D\,e^{(-i-1) \,Lk_0/\sqrt{2}} = \tilde{t}_{eh};\\
    A\,(i  - 1)\,e^{(i  - 1) \,Lk_0/\sqrt{2}}+ B\,(-i  + 1)\,e^{(-i  + 1) \,Lk_0/\sqrt{2}}+C\,(i  +1)\,e^{(i  +1) \,Lk_0/\sqrt{2}} + D\, (-i-1)\,e^{(-i -1) \,Lk_0/\sqrt{2}} = -i \sqrt{2\tilde\mu_\perp} \tilde{t}_{eh}.
\end{gather}
\end{widetext}
From these relations one can obtain analytic formulas for the transmission amplitudes:
\begin{widetext}
\begin{gather}
    \tilde t_{ee}= -\frac{\splitfrac{\Big\{(1+i) \sqrt{ \tilde\mu_\perp } e^{\frac{(1+i) Lk_0}{\sqrt{2}}}  (-\sqrt{2}  \tilde\mu_\perp +(2-2 i) \sqrt{ \tilde\mu_\perp }+ (\sqrt{2}  \tilde\mu_\perp +(2-2 i) \sqrt{ \tilde\mu_\perp }-i \sqrt{2} ) e^{(1+i) \sqrt{2} Lk_0}}{+ (i \sqrt{2}  \tilde\mu_\perp +(2-2 i) \sqrt{ \tilde\mu_\perp }-\sqrt{2} ) e^{\sqrt{2} Lk_0}+ (-i \sqrt{2}  \tilde\mu_\perp +(2-2 i) \sqrt{ \tilde\mu_\perp }+\sqrt{2} ) e^{i \sqrt{2} Lk_0}+i \sqrt{2} )\Big\}}}{( \tilde\mu_\perp -1)^2 e^{\sqrt{2} Lk_0}+( \tilde\mu_\perp -1)^2 e^{(1+2 i) \sqrt{2} Lk_0}-( \tilde\mu_\perp +1)^2 e^{i \sqrt{2} Lk_0}-( \tilde\mu_\perp +1)^2 e^{(2+i) \sqrt{2} Lk_0}-8  \tilde\mu_\perp  e^{(1+i) \sqrt{2} Lk_0}};\\
    \tilde t_{eh}= -\frac{2 \left(\sqrt{2}-(1-i) \sqrt{\tilde\mu_\perp }\right) \sqrt{\tilde\mu_\perp } e^{\frac{(1+i) Lk_0}{\sqrt{2}}} \left(-\tilde\mu_\perp +(-1-i \tilde\mu_\perp ) e^{\sqrt{2} Lk_0}+(1+i \tilde\mu_\perp ) e^{i \sqrt{2} Lk_0}+(\tilde\mu_\perp +i) e^{(1+i) \sqrt{2} Lk_0}-i\right)}{\left(\sqrt{2} \sqrt{\tilde\mu_\perp }+(-1-i)\right) \left((\tilde\mu_\perp -1)^2 e^{\sqrt{2} Lk_0}+(\tilde\mu_\perp -1)^2 e^{(1+2 i) \sqrt{2} Lk_0}-(\tilde\mu_\perp +1)^2 e^{i \sqrt{2} Lk_0}-(\tilde\mu_\perp +1)^2 e^{(2+i) \sqrt{2} Lk_0}-8 \tilde\mu_\perp  e^{(1+i) \sqrt{2} Lk_0}\right)}.
\end{gather}
\end{widetext}

% \section{Continuity of Currents in the NSN Structure}

% One can use the unitary condition of the scattering matrices to prove that in the NSN structure the heat does not accumulate in the superconducting region:
% %
% \begin{multline}
% \label{hc_conservation}
%     I^{L}_Q + I^{R}_Q = \frac{2}{h} \int dE\, E\, \{\delta f_L\,[1 - R^L_{ee} - R^L_{eh} - T^L_{ee} - T^L_{eh}]\\
%     +\delta f_R\,[1 - R^R_{ee} - R^R_{eh} - T^R_{ee} - T^R_{eh}]\} = 0.
% \end{multline}
% %
% This contrasts with the electric charge which does accumulate in the superconductor,
% %
% \begin{multline}
% \label{ec_conservation}
%     I^{L}_e + I^{R}_e = \frac{4e}{h} \int dE\, \{\delta f_L\,[R^L_{eh} + T^L_{eh}]
%     +\delta f_R\,[R^R_{eh} + T^R_{eh}]\}.
% \end{multline}
% %
\section{NXSXN Transmission Probabilities}
\label{Appendix2}
The superconducting part of a hybrid NXSXN structure is characterized by its own transmission and reflection amplitudes, which are given by
\begin{gather}
    t_{ee(hh)}=\frac{e^{\pm i p L}\sin{\alpha}}{\sin{(\alpha-i q L)}};\\
     r_{eh(he)}=\frac{\sinh{q L}}{i \sin{(\alpha-i q L)}}.
\end{gather}

Each X-part (e.g., quantum dot), can be simulated by a double barrier, which in turn is equivalent to a Fabry-Per\'ot interferometer.
Let us suppose that the inner (outer) barrier of such structure is described by the transmission $t_{i(o)}$ and reflection $r_{i(o)}$ coefficients.
Then the coefficients for each X-part are:
\begin{gather}
    \label{FP_t}
    t_{L(R)}^{e(h)} = t_i\,t_o\,e^{i\,k_{e(h)}\,d_{L(R)}}
    /(1-r_i\,r_o\,e^{2\,i\,k_{e(h)}\,d_{L(R)}});
    \\
    \label{FP_r}
    r_{Li(Ri)}^{e(h)} = r_i + r_o\,t_i^2\,e^{2\,i\,k_{e(h)}\,d_{L(R)}}
    /(1-r_i\,r_o\,e^{2\,i\,k_{e(h)}\,d_{L(R)}});
\end{gather}
where $k_{e(h)}$ is the electron's (hole's) wave vector inside the double barrier and $d_{L(R)}$ is the length of the left (right) double barrier.
If we apply the Breit-Wigner approximation\,\cite{BW} to $T^X=\big|t_{L(R)}^{e(h)}\big|^2$, we arrive at Eq.\,(\ref{lorenz}) describing transmission probability near the resonance.
Using Eqs.\,(\ref{FP_t}) and (\ref{FP_r}), we can calculate the transmission coefficients of the whole XSX structure:
\begin{gather}
    t^{\text{XSX}}_{eh}=t^e_L\,[t_{ee}\,r^e_{Ri}\,r_{eh} + r_{eh}\,r^h_{Li}\,t_{hh}]\,t^h_R/\mathcal{D};\\
    t^{\text{XSX}}_{ee}=t^e_L\,[t_{ee}\,(1-t_{hh}^2\,r^h_{Li}\,r^h_{Ri})+r_{eh}\,r^h_{Li}\,t_{hh}\,r^h_{Ri}\,r_{he}]\,t^e_R/\mathcal{D};
\end{gather}
where $\mathcal{D}$ is determined by multiple reflections inside the XSX structure:
\begin{multline*}
    \mathcal{D}=1-t^2_{ee}\,r^e_{Li}\,r^e_{Ri} - t^2_{hh}\,r^h_{Li}\,r^h_{Ri} - r_{eh}\,r_{he}\,(r^e_{Li}\,r^h_{Li} + r^e_{Ri}\,r^h_{Ri})\\
    + (t_{ee}\,t_{hh}-r_{eh}\,r_{he})^2\,r^e_{Li}\,r^e_{Ri}\,r^h_{Li}\,r^h_{Ri}.
\end{multline*}
%
% The analysis shows\,\cite{Sadovskyy::2015} that $T^\text{XSX}_{eh}=|t_{eh}|^2$ has its greatest value when $p L + \Big[\mathrm{arg}\,\Big(r^e_{Ri}\Big) - \mathrm{arg}\,\Big(r^h_{Li}\Big)\Big]\Big/2 = \pi n$, which was assumed in Fig.\,\ref{heat_switch}(c).

\end{document}